\documentclass[12pt]{article}
\def\BibTeX{{\rm B\kern-.05em{\sc i\kern-.025em b}\kern-.08em
   T\kern-.1667em\lower.7ex\hbox{E}\kern-.125emX}}
\usepackage{graphicx}
\usepackage{amsfonts}
\usepackage{latexsym}
\usepackage{epsfig}

\begin{document}

\title{Comments on ``Resource placement in Cartesian product of networks"  
[Imani, Sarbazi-Azad and Zomaya, J.\ Parallel Distrib.\ Comput.\ 70 (2010) 481-495]}
\author{Pranava K.\ Jha \\
\mbox{} \\
Dept.\ of Computer Science \\
St.\ Cloud State University \\
720 Fourth Ave.\ S. \\
St.\ Cloud, MN 56301-4498 \\
\mbox{} \\
pkjha@stcloudstate.edu
}

\date{January 2013}

\maketitle

%\newpage

%\thispagestyle{empty}
\noindent
{\bf Abstract:} The present note points out a number of errors, omissions, redundancies and 
arbitrary deviations from the standard terminology in the paper ``Resource placement in Cartesian 
product of networks," by N.\ Imani, H.\ Sarbazi-Azad and A.Y.\ Zomaya
[J.\ Parallel Distrib.\ Comput.\ 70 (2010) 481-495].

\bigskip

\noindent
{\bf Key words:} Critique, comments, resource placement, Cartesian product, graphs and networks.

\newpage
\section{Introduction} The present note is a critique of the paper ``Resource placement in 
Cartesian product of networks" by Imani, Sarbazi-Azad and Zomaya \cite{ISZ}. 
The paper under review contains a number of errors, omissions, redundancies and arbitrary 
deviations from the standard terminology. In particular, the authors (a) present certain results
that have existed in the literature for ages, (b) do not cite relevant references, 
(c) indulge in unnecessary redundancies, and (d) make statements that are vague, meaningless or
incorrect. In addition, there are a number of grammatical errors (not covered in the present
study).
%Meanwhile this note is not comprehensive with respect to the mistakes in the paper. 

In what follows, there is a section-by-section commentary on the paper under review.

% \section{Comment on Section 1} \label{sec-1}
% \begin{enumerate}
% \item Last sentence of the second paragraph (left column), p.\ 482: {\em To the best of
% our knowledge, this study is the first attempt to consider resource placement in product networks.} 
% \end{enumerate}

% This claim is patently incorrect.

% V.G.\ Vizing, The Cartesian product of graphs, Vyc. Sis., 9 (1963) 30-43.

% P. Ramanathan and S. Chalasani. Resource placement with
% multiple adjacency constraints in k-ary n-cubes. IEEE Trans.
% Parallel and Distributed Systems, 6(5):511–519,May 1995.

% H.-L. Chen and N.-F. Tzeng. Efficient resource placement
% in hypercubes using multiple-adjacency codes. IEEE Trans.
% Comput., 43(1):23–33, Jan. 1994.

% Myung M. Bae and Bella Bose, "Resource Placement in Torus-Based Networks"
% IEEE Transactions on Computers  
% Volume 46 Issue 10, October 1997 Page 1083-1092 
 
% K.W. Cattermole, "Communication networks based on the product graph"
% PROC. IEE, Vol. 124, No. 1, JANUARY 1977.

\section{Comments on Section 2} \label{sec-2}
\begin{enumerate}
\item In the fourth paragraph from the bottom (left column, p.\ 482), the authors refer to a paper 
by Alrabady et al \cite{AMC} for certain definitions of resource allocation strategies. Interestingly 
enough, the forenamed paper itself contains several errors. For example, 
\begin{enumerate}
\item Lemma 1 \cite{AMC} (p.\ 62): ``Any perfect resource set is a dominating set and a maximal
independent set." The plain truth is that a maximal independent set is necessarily a dominating set.
\item At the top of the second column \cite{AMC} (p.\ 62): ``$\cdots$ looking for a dominating set 
 and a maximal independent set is an NP-complete problem," which is far from true. Indeed, obtaining 
a smallest dominating set, or a smallest/largest maximal independent set is NP-complete. 
\end{enumerate}
\item In the third paragraph from the bottom (left column, p.\ 482), replace ``less than or equal to
$d$" by ``greater than $d$."
\item On the sixth line from the bottom (left column, p.\ 482), replace ``$d=1$ and $m=1$" by
``$m=1$ and $d=1$."
\item Definition 2 (right column, p. 482) deals with $Vol_G(d, c)$ that is the number of vertices 
within a distance of 
$d$ from a fixed vertex $c$ in $G$. In general, $Vol_G(d, c_1)$ need not be equal to $Vol_G(d, c_2)$ 
for $c_1 \neq c_2$, yet the authors make use of $Vol_G(d, c)$ in the statement and proof of 
Theorem 1 without ever referring to vertex $c$. They implicitly assume that $Vol_G(d, c)$ is 
independent of $c$, which need not be true with respect to an arbitrary graph $G$.
Note that the statement of Theorem 1 starts with ``For any graph $G=(V,E)\cdots$."
\item The foregoing concept of {\em volume} is not used anywhere in Sections 4 and 5, which constitute
the main body of the paper, so the definition itself is useless.
\item On p.\ 482 (second column), the authors introduce the term ``homogeneous" in the sense of 
``isomorphic." When there exists a world-wide unanimity on the concept 
of isomorphism, introducing a different term for that purpose is patently incorrect. 
Strangely enough,  the authors themselves,  in an earlier paper \cite{ISZM}, employed the terms 
``isomorphic" and ``isomorphism" in the usual sense. Moreover, 
there already exists the concept of a homogeneous graph in the literature that is completely 
different from that of isomorphism: {\em A graph $G$ is said 
to be homogeneous if for any two isomorphic vertex-induced subgraphs $\langle X\rangle$ and 
$\langle Y\rangle$ of $G$, there exists some isomorphism between $\langle X\rangle$ and 
$\langle Y\rangle$ that extends to an automorphism of $G$ \cite{Ga}}.
\item Corollary 1 (p.\ 482) deals with the vertex partition of a Cartesian product of several graphs into 
subgraphs isomorphic to a fixed factor graph. A detailed proof without any citation purports 
that this is authors' original idea. However, this property (and the related concept of projection on
a fixed co-ordinate) of a Cartesian-product graph has existed in the literature for a long time 
\cite{Sa, Vi}. In particular, it was illustrated in the book on product graphs by Imrich and 
Klav\v zar \cite{IK} (pp. 30-31), which was published more than ten years before the 
publication of the paper under review, yet the authors do not cite that book or any other source
for that purpose. Interestingly enough, they cite a paper jointly by Klav\v zar \cite{KY} that
itself refers to that book. 
\end{enumerate}

\section{Comments on Section 3} \label{sec-3}
\begin{enumerate}
\item Definition 3 (p.\ 483): Replace ``$0\le i\le |Q_G|$" by ``$1\le i\le |Q_G|$". 
\item Proof of Corollary 2 (p.\ 483): 
\begin{enumerate}
\item Replace $Q_G(k, d)$ by $Q_G(m, d)$. 
\item Replace ``$R$ contains" by ``each $R_i$ contains".
\end{enumerate}
\item The proof of Corollary 3 (p.\ 483) is completely redundant since the statement is an obvious
consequence of Definition 3 and Corollary 2. Likewise Corollary 4 (p.\ 483) follows from the fact that 
$Q_G$ constitutes a vertex partition of $G$, so its proof is equally redundant.
\end{enumerate}

\section{Comments on Section 4} \label{sec-4}
\begin{enumerate}
\item On p. 484, the authors present Algorithm 2-{\em HMP}, which consists of steps (a) and (b).
Immediately thereafter, they prove in Theorem 2 that Step (b) of that algorithm is redundant. An 
identical situation arises on p. 485 with respect to Algorithm 2-{\em HTP} and Theorem 3, respectively. 
This kind of baroque has no place in a journal where space is at a premium. The authors have a 
responsibility to present an algorithm succinctly, so there is no chaff around it.
\item In the proof of Theorem 4 (p.\ 485), the authors write, `` $\cdots$ 
a bijective function, i.e., a function that is both surjective and injective." There is absolutely no 
need to educate the reader of a premium journal that a bijective function is both surjective and injective.  
\item In the proof of Theorem 4 (p.\ 485), the authors write, `` $\cdots$ the output of our algorithm
is a $|Q_H| = |Q_{G_1}| = |Q_{G_2}|$ cubic matrix $M$," which does not make sense at all. 
Indeed, there is no ``cubic matrix" anywhere else in the paper. A little later, 
they present Algorithm {\em DM-MF} (that is a part of the proof of the same theorem) in which $M$ appears 
as the matrix $M_{|Q_H|\times |Q_H|}$. Further, at Step (5) of the same algorithm (p.\ 486), $M$ appears as
the matrix $M_{a\times b}$ without any subsequent discussion on how $a$ and $b$ are related to $|Q_H|$.
\item In the statements of Theorems 5 and 6 (pp.\ 486-487), the authors start with the hypothesis
that $G_1$ and $G_2$ are arbitrary graphs, and then immediately impose the condition that 
$|Q_{G_1}| = |Q_{G_2}|$ where, in addition, there exists a bijection $\phi_1$ from  
$Q_{G_1}$ to $Q_{G_2}$. The condition is severe, hence at odds with the premise that  
$G_1$ and $G_2$ are arbitrary graphs.
\item In the first paragraph of the proof of Theorem 5 (p.\ 486), $v\in R_{1,1}$ as well as
$v\in R_{H,1}$, where $R_{1,1}\in Q_{G_1}$ and $R_{H,1}\in Q_{G_1\times G_2}$. It is impossible
to reconcile the membership of $v$ in both $R_{1,1}$ and $R_{H,1}$.
\item p.\ 487, second column: Replace ``$Q_{G_i} = Q_{G_j}$, $1\le i, \, j\le k$, $i\ne j$" by
``$|Q_{G_i}| = |Q_{G_j}|$, $1\le i, \, j\le k$".
\item Step (b) of Algorithm {\em IHTP}($H$) on p.\ 488 is as redundant as the respective step in each of
Algorithm 2-{\em HMP} (p.\ 484) and Algorithm 2-{\em HTP} (p.\ 485).
\item In the proof of Theorem 9 (p.\ 488), ``$\bigcup_{i=1}^k V(G'_i) = u$" must be replaced by
``$\bigcap_{i=1}^k V(G'_i) = \{u\}$".
\item Proof of Corollary 5 (p.\ 488) is trivial, hence unnecessary.
\end{enumerate}

\section{Comments on Section 5} \label{sec-5}
\begin{enumerate}
\item On the fourth line in the paragraph after Algorithm {\em SDP} (p.\ 489): Replace ``an arbitrary
graph in $H$" by ``an arbitrary node in $H$".
\item In the paragraph immediately above Corollary 6 (p.\ 489): Replace ``A known 2 distance-2 
placement for $C_6$ is assumed" by ``A known one-perfect distance-one placement for $C_6$ is 
assumed".
 
\end{enumerate}


\baselineskip 12pt
\begin{thebibliography}{99}
\bibitem{AMC} A.I.\ Alrabady, S.M.\ Mahmud and V.\ Chaudhary, Placement of resources in the
star network, Proc. IEEE International Conference on Algorithms and Architectures for 
Parallel Processing, 1996.
\bibitem{Ga} A.\ Gardiner, Homogeneous graphs, J. Comb. Theory, Ser.\ B 20 (1) (1976) 94-102.
\bibitem{ISZ} N.\ Imani, H.\ Sarbazi-Azad and A.Y.\ Zomaya, Resource placement in Cartesian product
of networks, J.\ Parallel Distrib.\ Comput.\ 70 (2010) 481-495.
\bibitem{ISZM} N.\ Imani, H.\ Sarbazi-Azad, A.Y.\ Zomaya and P.\ Moinzadeh, Detecting threats in star 
graphs, IEEE Trans. Parallel Dist.\ Syst.\ 20 (2009) 474-483.
\bibitem{IK} W.\ Imrich and S.\ Klav\v zar,  Product Graphs: Structure
and Recognition, John Wiley \& Sons, New York, NY, 2000.
\bibitem{KY} S.\ Klav\v zar and H.-G.\ Yeh, On the fractional chromatic number, the chromatic number, 
and graph products, Discrete Math.\ 247 (2002) 235-242.
\bibitem{Sa} G.\ Sabidussi, Graph multiplication, Math.\ Z.\ 72 (1960) 446-457.
\bibitem{Vi} V.G.\ Vizing, The Cartesian product of graphs (Russian), Vyc.\ Sis.\ 9 (1963) 30-43. 
English translation in: Comp.\ El.\ Syst.\ 2 (1966) 352-365.

\end{thebibliography}
\end{document}